\documentclass[twocolumn,showpacs,nofootinbib,prl,reprint]{revtex4}
\usepackage{graphicx}

\newcommand{\del}{\partial}

\newcommand{\eps}{\epsilon}

\newcommand{\bold}[1]{{\mbox{\boldmath $#1$}}}    
\newcommand{\rr}{\bold{r}}			
\newcommand{\p}{\bold{p}}                       
\newcommand{\kk}{{\bf k}}			
\newcommand{\bfv}{\bold{v}}                    
\newcommand{\etal}{{\em et al.}}                
\newcommand{\beq}{\begin{equation}}
\newcommand{\eeq}{\end{equation}}
\newcommand{\beqar}{\begin{eqnarray}}
\newcommand{\eeqar}{\end{eqnarray}}
\newcommand{\bd}{\begin{itemize}} 
\newcommand{\ed}{\end{itemize}} 
\newcommand{\bc}{\begin{center}}
\newcommand{\ec}{\end{center}}

\newcommand{\be}{\begin{equation}}
\newcommand{\ee}{\end{equation}}
\newcommand{\ba}{\begin{array}}
\newcommand{\ea}{\end{array}}
\newcommand{\bfig}{\begin{figure}}
\newcommand{\efig}{\end{figure}}

\begin{document}

\title{Fluctuations and Symmetry Energy in Nuclear Fragmentation Dynamics}
\author{M. Colonna}
\affiliation{
        INFN-Laboratori Nazionali del Sud, I-95125, Catania, Italy\\}

\begin{abstract}
Within a dynamical description of nuclear fragmentation, based on the
liquid-gas phase transition scenario, we explore the 
relation between neutron-proton density fluctuations and nuclear 
symmetry energy.
We show that, along the fragmentation path,  
isovector fluctuations 
follow the evolution of the local density and approach an equilibrium value connected 
to the local symmetry energy.  
Higher density regions are characterized by smaller average asymmetry
and narrower isotopic distributions.
This dynamical analysis points out  that  
fragment final state isospin fluctuations
can probe the symmetry energy of the density domains from which fragments originate.   


\end{abstract}
\maketitle

The dynamics and thermodynamics of complex systems present 
general aspects, of interest in different domains of 
physics. A rather important issue is the identification 
of the occurrence of phase transitions. This is relevant for many
microscopic or mesoscopic systems, from metallic clusters to 
Bose condensates and nuclei 
\cite{Bose,metalli,Moretto}.
In particular, the analysis of two-component systems has recently evidenced
new interesting features \cite{Aug,Cha,Ciccio_spin,Muel}.  

Under suitable conditions of density and temperature, the nuclear 
Equation of State (EoS) foresees the possibility 
of phase transitions from the liquid to the vapour phases,
a scenario often evoked to explain the multifragmentation phenomenon \cite{bert,bow,report_borderie}. 
As a consequence of the two-component structure of nuclear matter, 
constitued by protons
and neutrons, a crucial role is  
played by the low-density behavior of the isovector part of the interaction and the corresponding term in the 
nuclear EoS, the symmetry energy \cite{rep},
on which many investigations are concentrated \cite{betty,shetty,francesca,galichet,Chimera}.      
We stress that  this information is essential 
in the astrophysical context, for the understanding of the properties of
compact objects such as neutron stars, which crust behaves as low-density 
asymmetric nuclear matter \cite{Lattimer,Duc}.
Moreover, the density dependence of the symmetry energy affects
the structure of exotic nuclei and 
the appearance of new features involving the neutron skin \cite{Colo}.

A connection between the characteristics 
of clusters emerging 
from nuclear fragmentation 
and the symmetry energy has been proposed, in the framework of macroscopic statistical models \cite{bot,
tsang,lefevre}.
However it would be important to explore this issue 
within a full dynamical description of the fragmentation process. 
Here we undertake such a kind of study for systems facing
low-density (spinodal) 
instabilities  and first-order phase 
transitions \cite{rep}. 
We investigate the coupling between the development of neutron-proton density 
fluctuations (isovector fluctuations), 
to which isotopic properties are connected, 
and the growth of unstable modes of the total density, leading to the formation of nuclear
drops (fragments).  
Thus the aim  of this work is to examine the behavior of isovector fluctuations in
rapidly evolving systems, 
to probe their possible
relation to the symmetry energy and its density dependence.

Theoretically the evolution of complex systems  
can be described
by a one-body transport equation with a fluctuating term, that incorporates the
effects of the unknown many-body correlations, 
the so-called
Boltzmann-Langevin equation (BLE) \cite{Ayik,BL}. 
We follow the approximate treatment to the BLE 
presented in Ref.\cite{Salvo}, 
the Stochastic Mean Field (SMF) model.
We solve the following equation for the time evolution of
the semiclassical one-body distribution function $f(\rr,\p,t)$:
\beq\label{BL_ext}
{\del f\over\del t}
+\bfv\cdot{\del f\over\del\rr}
-{\del U\over\del\rr}\cdot{\del f\over\del\p}=
\bar{I}_{coll}[f]+{\del U_{ext}\over\del\rr}\cdot{\del f\over\del\p},
\eeq
where $U[\rho]$ is the self-consistent mean-field potential,
$\bar{I}_{coll}[f]$ is the average 
collision integral and $U_{ext}(\rr)$ represents an external, stochastic field.
The coordinates of isospin are not shown for brevity.  
Within such a framework, 
the effective nuclear potential $U$ is derived from energy functionals that 
usually contain a
term proportional to $I^2$, the  symmetry energy  $E_{sym}(\rho,I)/A \equiv C_{sym}(\rho)I^2$ (with $I\equiv (\rho_n-\rho_p)/\rho$ and $\rho$, $\rho_n$, $\rho_p$ denoting 
total, neutron and proton densities, respectively).



Let us consider the behavior of nuclear matter
prepared with a uniform density distribution $\rho_0$
and with a Fermi-Dirac momentum distribution 
corresponding to a specified temperature $T$. 
The system is confined within a cubic box, with side $L = 19~fm$, 
with periodic boundary conditions imposed.
The linear response analysis allows one to get a first insight into the fluctuation dynamics. 
For two-component matter one can identify two types of independent
modes of the phase-space density: isoscalar-like modes, where neutrons and protons oscillate in phase,
and isovector-like modes, with neutrons and protons oscillating out-of phase. In particular, in 
the case of symmetric nuclear matter, the two types of modes correspond to oscillations of 
$f^s = f_n + f_p$ (isoscalar modes) and of 
$f^v = f_n - 
f_p$ (isovector modes). 
Let us denote by $f^q_{\kk}(\p,t)$ ($q=s,v$) the Fourier transform, with respect to $\rr$, of the difference
$\delta f^q=f^q-f^q_0$, where $f^q_0$ is the system initial phase-space density.
The equation of motion for these Fourier coefficients 
follows readily from (\ref{BL_ext}),
\begin{equation}
\label{EoM}
{\del\over\del t}f^q_{\kk} +i\kk\cdot\bfv f^q_{\kk}
-i{\del U^q_k\over\del\rho^q} \kk\cdot\bfv {\del f_0\over\del\eps}\rho^q_{\kk}
= i{\cal F}^q_\kk \kk\cdot\bfv {\del f_0\over\del\eps}~,
\end{equation}
Here $\del U^q_k/\del\rho^q$ represents the appropriate Fourier component
of the derivative of the effective field with respect to the density $\rho^q$ 
and ${\cal F}^q_\kk(t)$ is the Fourier component
of the external field. 
Furthermore,
$\rho^q_\kk(t)$
is the Fourier transform of the density fluctuation 
$\delta\rho^q(\rr)$. 
Finally, since we will restrict our analysis to rather low temperatures,
in Eq.(2) we have ignored the average collision term $\bar I_{coll}$,
since its effect is relatively small \cite{Pethick}.


\begin{figure}
\includegraphics[width=8.5cm]{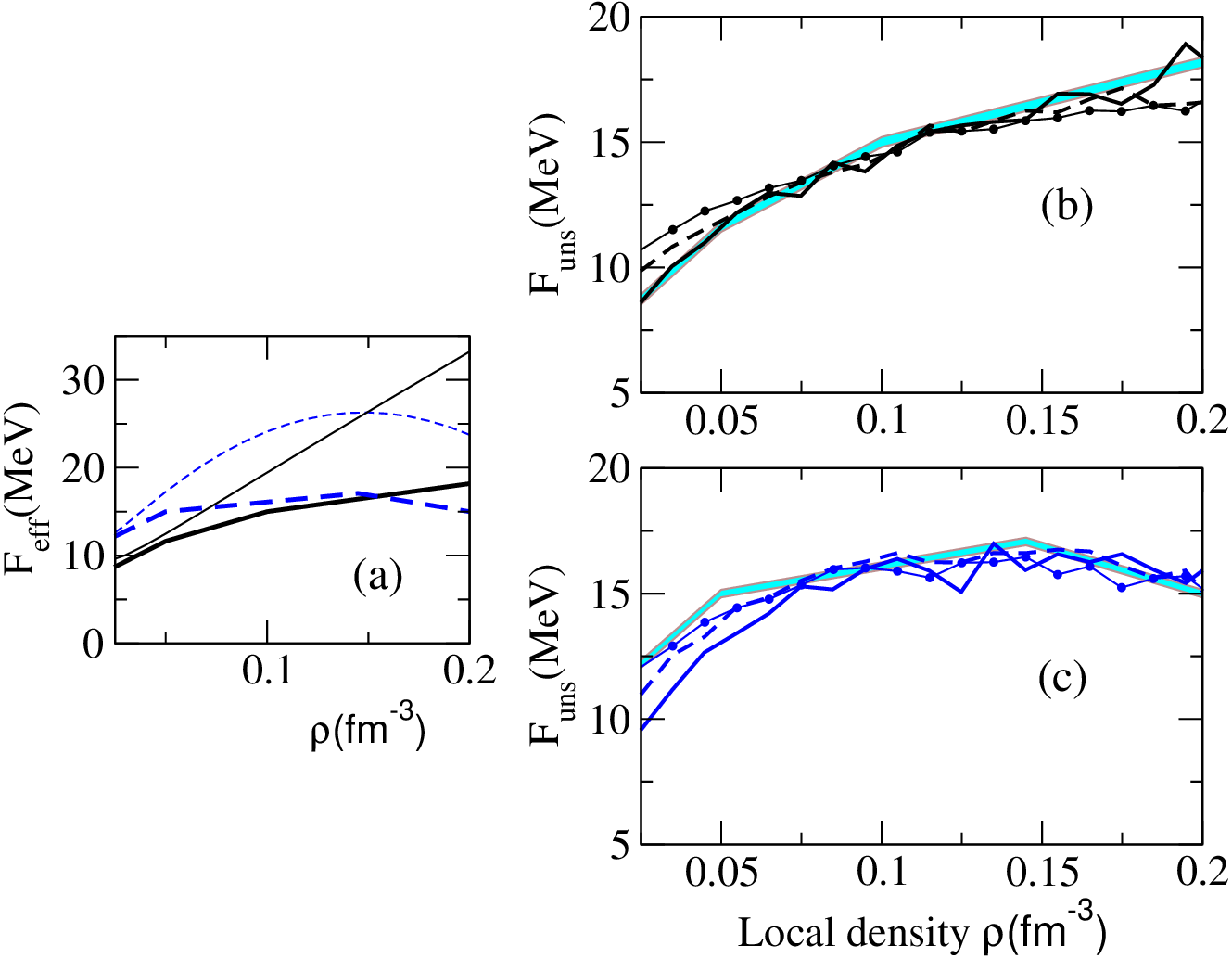}

\caption{
Panel (a): 
The quantity $F^v_{eff}$, extracted from SMF simulations, for
stable nuclear matter in several density conditions and at  
temperature T = 3 MeV (thick lines).  Thin lines show the density dependence of the symmetry free energy 
${\bf F}_{sym}$ .  Full lines: asy-stiff EoS. Dashed lines: asy-soft EoS.
Panel (b):
The quantity $F^v_{uns}$, evaluated at the freeze-out time, as a function of the local density
for unstable systems with 
initial density $\rho_1$ (full lines), $\rho_2$ (dashed lines), $\rho_3$ 
(dotted lines). Asy-Stiff EoS.
Panel (c): The same as in panel (b), for asy-soft EoS. In panels (b) and (c)
thick gray (cyan) lines represent the same results shown in panel (a) as thick lines. 
}

\end{figure}

For stable modes, 
the equilibrium variance $\sigma^q_k$ associated with the 
fluctuation 
$\rho^q_\kk$ 
is linked to the physical quantities 
that 
characterize the response of the system to the action of the external force ${\cal F}^q_\kk$, see Eq.(2).
According to the  fluctuation-dissipation theorem \cite{Landau}, one can write:
$\sigma^q_k = {T /{F^q(k)}}$,
where $F^q(k) = ({\del U^q_k\over\del\rho^q} + 1/{\cal N})$, with ${\cal N} = 
-{4\over (2\pi)^{3}}\int d\p {\del f_0\over\del\eps}$. 
We notice that $F^q$ is nothing but the second derivative of 
the system free energy
density with respect to the density $\rho^q$.
Considering the inverse Fourier transform of $\rho^q_\kk$
we obtain, for the equilibrium spatial density correlations, in a cell of
volume $\Delta V$:
$$
\sigma_{\rho^q}^{(eq)}(\Delta V) \equiv <\delta\rho^q(\rr) \delta\rho^q(\rr)> =  
$$
\beq
{1\over{(2\pi)^3}}\sum_\kk \sigma^q_k ~ d\kk = {T
\over{\Delta V}}  <1/F^q(k)>_\kk,
\eeq
where the average extends over all $\kk$ modes.  



Focusing on isovector modes, 
the potential $U^v_k$ represents the Fourier 
transform of the symmetry potential
$U_{sym}[\rho_0(\rr)] = 2 {\rho_v \over \rho_0} 
\int d\rr'~ C^{pot}_{sym}[\rho_0(\rr')]\cdot g_\sigma(|\rr-\rr'|)$, where  $C^{pot}_{sym}$ denotes the potential part of  
$C_{sym}$
and the smearing function $g_\sigma$ is introduced 
to account for the finite range of the nuclear interaction. 
Thus we obtain:  
$F^v(k) =  2C_{sym}^{pot}(\rho_0)~ g_\sigma(k)/\rho_0 + 1/{\cal N} \equiv
2F_{sym}(k)/\rho_0$. 
We note that the function ${\bf F}_{sym}(\rho_0) \equiv F_{sym}(k=0)$ simply coincides with the volume symmetry free energy, that 
at zero temperature 
reduces to the symmetry energy
$C_{sym}(\rho_0)$. 
We can write:
\beq
 <1/F^v(k)>_\kk = {{\rho_0 }\over {2}}  <1/F_{sym}(k)>_\kk \equiv    {{\rho_0 }\over {2F^v_{eff}}}.
\eeq
Hence we find that equilibrium fluctuations of the isovector density can be connected to 
an ``effective'' symmetry free energy $F_{eff}^v$
that, owing to the $k$ dependence of the symmetry potential,  is smaller than the 
free energy ${\bf F}_{sym}$. 
  
In asymmetric matter, the findings discussed above still hold for isoscalar-like and 
isovector-like oscillations.  

Now let us go back to the full non-linear equations (1), that are solved numerically
with the test particle method \cite{BaranNPA703}.
We have performed SMF calculations for nuclear matter  prepared at 
initial temperature T = 3 MeV and
in several density conditions. 
Here we also take account of fluctuations in the isovector channel, which
were neglected in Refs.\cite{BaranNPA703,rep1}.
Isovector fluctuations can be extracted from
the model by simply rescaling the variance by the number of test particles employed
in the simulation \cite{alfio}.

We adopt momentum-independent effective interactions
corresponding to a soft EoS, with compressibility modulus $K = 200~ MeV$. 
The coefficient $C_{sym}$ gets a
kinetic contribution just from basic Pauli correlations and a potential
part,  $C^{pot}_{sym}$,  from the 
isospin dependence of the interaction. 
For the local density  ($\rho$) dependence of  $C_{sym}^{pot}$
we consider two representative parametrizations: 
one with a linearly increasing behaviour   
 with density (asy-stiff),  $C_{sym}^{pot}(\rho) = 90~\rho$ (MeV),
and one with a kind of saturation above normal
density (asy-soft),  $\displaystyle{C_{sym}^{pot}(\rho)}=\rho~(238-1009~\rho)$ (MeV) 
\cite{BaranNPA703,rep1}. 
We notice that 
at the temperature considered in the calculation,
which is within the typical range observed in multifragmentation 
\cite{report_borderie},
the symmetry energy $C_{sym}$  is very close to 
${\bf F}_{sym}$. 
As smearing function $g_\sigma$, we take a gaussian with width  $\sigma = 0.9~fm$.
With this choice, 
for nucler matter at saturation density ($\rho_0 = \rho_{sat}=0.145~fm^{-3}$),
Eq.(4) gives $F_{eff}^v = 0.7~ {\bf F}_{sym}$. 

Let us consider first, for the sake of simplicity, the case of symmetric matter ($I = 0$).
We first concentrate on  isovector fluctuations for 
uniform matter at rest, where equilibrium conditions are fulfilled.  
Thus, in order to avoid the development of volume instabilities
at  low density \cite{rep}, we switch-off
in the calculations the isoscalar part of the nuclear potential. 
Then we calculate the isovector fluctuation variance  
$\sigma_{\rho^v} = <(\delta\rho_n(\rr) - \delta\rho_p(\rr))^2>$, where the average is performed 
over cells of volume $\Delta V = 1~fm^3$.  
The effective symmetry free energy can be extracted from the numerical variance exploiting 
Eqs.(3,4).
This quantity 
is displayed in panel (a) of 
Fig.1  
as a function of the matter density, for 
the two parameterizations of the symmetry energy introduced above 
(thick lines, full for asy-stiff and dashed
for asy-soft), and compared with the corresponding symmetry free energy ${\bf F}_{sym}$. 
The numerical results generally go with 
the analytical estimation discussed above: 
Owing to the $k$ dependence of the symmetry potential, the extracted $F_{eff}^{v}$
is lower than the symmetry free energy, being reduced by about $30\%$ at saturation density,
and exhibits a density dependence connected
to the asy-stiffness of the effective interaction employed in the simulations. 
\begin{figure}
\includegraphics[width=6.5cm]{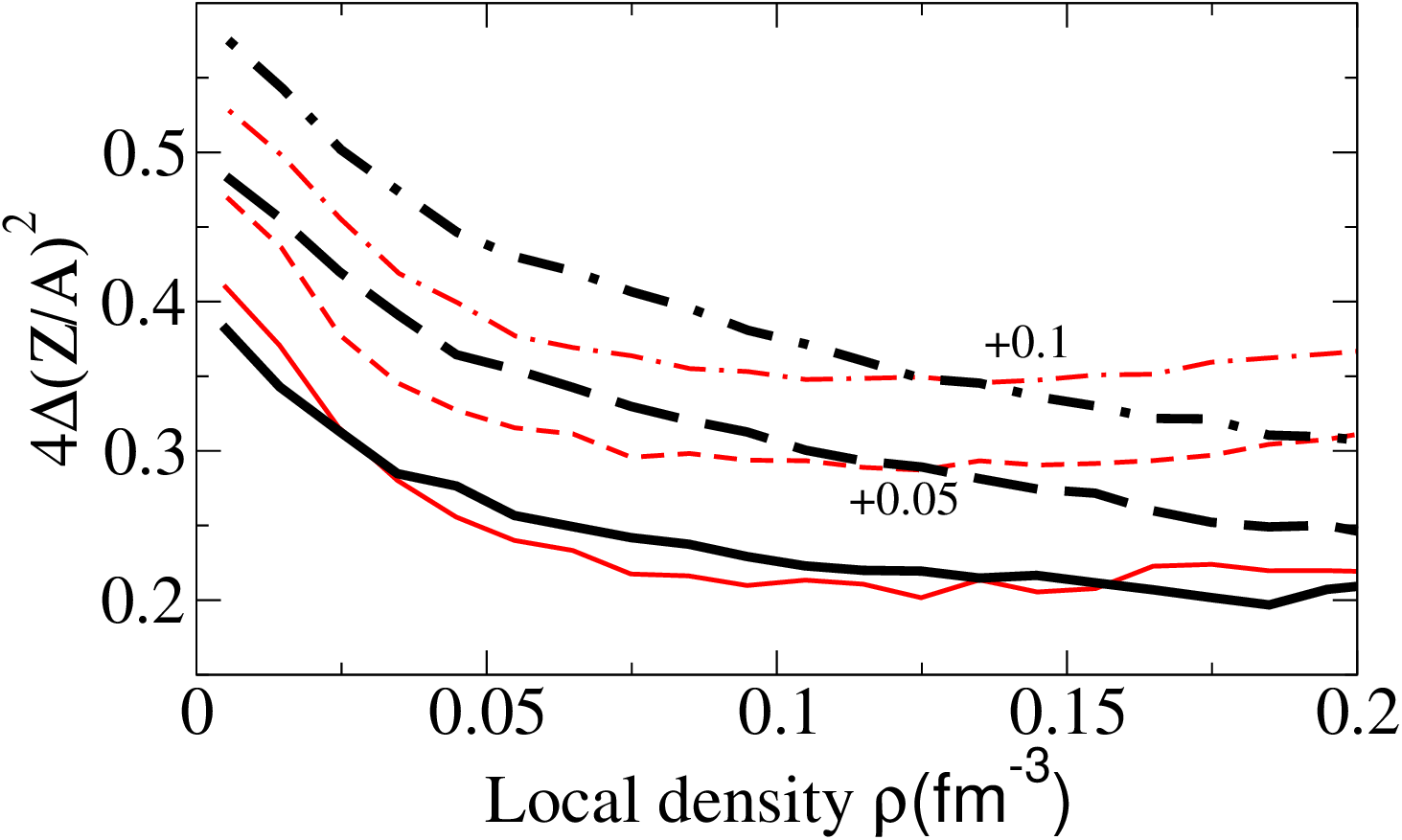}

\caption{(Color online) 
The quantity $4\Delta({\overline{Z/A}})^2)$ (see text) is plotted as a function of
the local density, for systems having  
initial density $\rho_1$ (full lines), $\rho_2$ (dashed lines), $\rho_3$ 
(dot-dashed lines).
Curves are shifted for a better visibility. 
Black, asy-stiff EoS; gray (red), asy-soft EoS.}
\end{figure}

The evaluation of the equilibrium isovector fluctuations of 
stable matter 
can be used as a benchmark for the 
general and more interesting case where unstable systems are let evolve. 
Calculations have been performed taking, as initial density $\rho_0$, 
three values inside the spinodal region: 
$\rho_1 = 0.0245~fm^{-3}$, 
$\rho_2 = 2\rho_1$ and $\rho_3 = 3\rho_1$.  Moreover, for each case, we have considered symmetric matter 
(system (1), $I_1 = 0$)
and asymmetric matter (system (2), $I_2 =0.142$). 


Now the system may develop density fluctuations, 
so locally the density gets larger (density bumps, leading to fragments) 
or smaller (vapour) than
the initial value \cite{rep}. 
The separation between the two regimes is smooth, so that the local density $\rho$ may vary between zero and
values around the saturation density.  
Our analysis is performed at the ``freeze-out'' time $t = 200~fm/c$, when isoscalar density 
fluctuations saturate. 
At this time, the average density of the regions
having $\rho$ larger than $\rho_0$ 
goes
from $0.064~fm^{-3}$ (in the $\rho_1$ case) to 
$0.10~fm^{-3}$ ($\rho_2$ case) and $0.12~fm^{-3}$ ($\rho_3$ case).

Our aim is to investigate the behavior of isovector fluctuations on the short time scale 
(the ``freeze-out'' time) associated with fragment formation.  
Isovector fluctuations are evaluated as a function of the local density inside the fragmenting system, 
looking at
the variance of the isovector density $\rho^v$ in cells having the same local density $\rho$.
As a measure of the isovector variance $\sigma_{\rho^v}$,  we consider the quantity $F^{v}_{uns} = 
(\rho~T) / (2\Delta V \sigma_{\rho^v})$, that coincides with  $F^{v}_{eff}$ if equilibrium is reached
(see Eqs.(3,4)). 
Results for $F^{v}_{uns}$, obtained in the case of symmetric matter,
 are displayed in Fig.1 as a function of the local density, 
for the three initial density values considered, see panels (b) and (c). 
Quite interestingly, 
isovector fluctuations
follow the local value of the symmetry energy 
independently of the initial conditions of the system. 
Indeed the three curves associated with the different initial densities (full, 
dashed and dotted lines for $\rho_1$, $\rho_2$ and $\rho_3$, respectively) are rather close to each other
and they are also close, 
for each given local density, to the equilibrium results discussed above
(here plotted as thick gray (cyan) lines), thus locally $F^{v}_{uns} \approx  F^{v}_{eff}$.
These results indicate that, as soon as density fluctuations 
start to develop, a quick rearrangement of 
isovector fluctuations takes place, so that the equilibrium value corresponding to the new 
actual local density is approached.
Indeed isovector-like oscillations are characterized by a much shorter time scale,
with respect to the growth of the unstable modes \cite{Ciccio}.
Thus important coupling effects between isoscalar and isovector oscillations are emerging from the
solution of the full non-linear Eqs.(1). 

Calculations have also been performed for the asymmetric system (2), leading
to results very close to the ones displayed in Fig.1.
In the latter case one can also discuss the isospin distillation mechanism, that induces a deviation of 
the local asymmetry from the system initial value \cite{rep1}. 
In particular, we consider the following
 density-dependent quantity,
derived from the symmetric system (1) and the asymmetric system (2): 
$\Delta({\overline{Z/A}})^2=({\overline{Z/A}})_1^2 - ({\overline{Z/A}})_2^2$, 
where $({\overline{Z/A}})_i$ (with $i=1,2$) represents, for the system (i) the average proton fraction
of cells having the same local density $\rho$.
This quantity is displayed in Fig.2 as a function of $\rho$.
The different curves correspond to the two EoS (gray (red) lines for soft, black lines for stiff) 
and the three initial densities considered. As a general trend, we observe the well known 
behavior of 
asymmetric systems: The low-density regions become more neutron rich, while high density regions are 
more symmetric, just in connection with the density dependence of the symmetry energy coefficient 
$C_{sym}(\rho)$. 
Here what is interesting to notice is that the distillation mechanism goes together with
the density-dependent behavior of the isovector variances described just above. 
As shown by Figs.1-2,  
large density domains are associated with 
larger $F^v_{eff}$ (i.e. smaller fluctuation width $\sigma_{\rho^v}/\rho$)
and smaller asymmetry, 
whereas low density regions are on average more asymmetric, but
also more fluctuating. 
\begin{figure}
\includegraphics[width=6.cm]{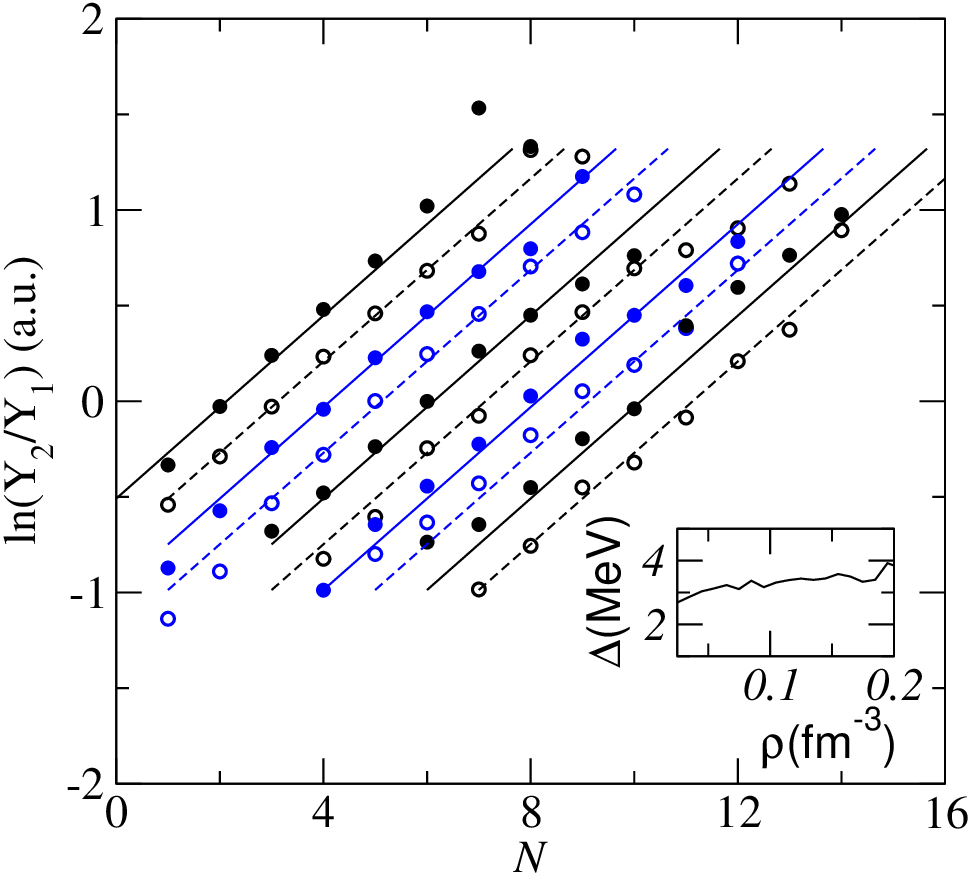}
\caption{(Color online) 
The quantity $ln(Y_2/Y_1)$ is plotted as a function of $N$, for the charges
$Z = 1-10$, in the case of the systems with initial density $\rho_1$.
The stiff parametrization in considered.  
Lines are to guide the eye.  
The inset shows the product 
 $\Delta = 4\Delta({\overline{Z/A}})^2 \cdot
F_{eff}^v$, as a function of the local density. 
} 
\end{figure}

Let us move to study 
the probability  $Y(Z,N)$ to find, inside a volume $V$, 
a given number of protons and neutrons,
$Z = {\rho_{n,V}}~V$ and $N = {\rho_{p,V}}~V$. 
${\rho_{n,V}}$ and  ${\rho_{p,V}}$ denote  neutron and proton densities
averaged over 
$V$, whose sum yields the density $\rho_V$.
Here we consider $V = (5.5~fm)^3$.   
The quantity $Y(Z,N)$ is proportional to 
the probability of
getting, in the volume $V$, 
a specific variation of the isovector density $\rho^v$, with respect to the average
${\overline{\rho^v}}$:
$P(\rho^v) \approx exp-(\rho^v - {\overline{\rho^v}})^2/(2\sigma_{\rho^v})$.
Using the identity $\rho^v/\rho_V = I = (N-Z)/A$ and 
considering the equilibrium amplitude of $\sigma_{\rho^v}$
(see Eqs.(3,4)), one can write, 
for the yield ratio between systems (1) and (2):
\beq
ln(Y_2/Y_1) \approx [(I-{\overline I}_1)^2 - (I-{\overline I}_2)^2]A~F_{eff}^{v}/T,
\eeq
where $F_{eff}^v$ 
and the average asymmetry  ${\overline I_{i}}$
are functions of 
${\rho}_V$, 
(in our case 
${\overline I_{1}}=0$).
We notice that the ratio $Y_2/Y_1$ does not depend explicitly on the volume $V$.
After some algebra, Eq.(5) can be rewritten as:
\beq
ln(Y_2/Y_1) \approx [({\overline I}^2_1 - {\overline I}^2_2)(N + Z) - 2({\overline I}_1-{\overline I}_2)
(N - Z)]F_{eff}^{v}/T,
\eeq
Expressing 
${\overline I_i}$ in terms of the average proton or neutron fraction, we finally get:
$ln(Y_2/Y_1) \approx \alpha N + \beta Z$, with:
$$
\alpha({\rho}_V) 
= 4\Delta({\overline{Z/A}})^2 F_{eff}^{v}/T,
$$
\begin{equation}
\beta({\rho}_V) 
= 4\Delta({\overline{N/A}})^2 F_{eff}^{v}/T.
\end{equation}
Thus   we recover the standard isoscaling relations \cite{bot}, but with density-dependent coefficients $\alpha({\rho}_V)$ and  $\beta({\rho}_V)$, linked to the effective symmetry free energy $F_{eff}^{v}$. 

The behavior of the exponent $\alpha$ is illustrated in Fig.3, where
we plot the quantity $ln(Y_2/Y_1)$ as a function of $N$, for the charges
$Z = 1-10$. Inspite of the implicit density dependence of the isoscaling parameters, 
we note that the slope $\alpha$ 
is the same for all charges. 
This result follows from the opposite trend, shown by Figs. 1-2, of the two quantities 
 $\Delta({\overline{Z/A}})^2$ and
$F_{eff}^v$, so that the
product keeps almost constant (see the inset of Fig.3).
More precisely, the quantities   $\Delta({\overline{Z/A}})^2$ and  $\Delta({\overline{N/A}})^2$
go approximately as 
${\rho_0\over \rho}~{{\partial C_{sym}}\over{\partial \rho}}|_{\rho = \rho_0}$ \cite{Ciccio},
counterbalancing 
the density dependence of $F_{eff}^v$. 
In the case of
a linear behavior of  
$F_{eff}^v$, i.e. close to the conditions of our stiff case,
the isoscaling parameters, Eqs.(7), would be exactly constant. 
However also in the soft case (not shown in the figure), the exponent $\alpha$ is roughly the
same for all Z values (within 7$\%$.)
Within our framework, 
the nearly constant value of $\alpha$ (or $\beta$) 
inside the fragmenting system
could be at the origin of
the experimental observation of the same isoscaling parameter for the several products 
issued from nuclear reactions \cite{tsang}, which in principle may originate from different
density regions and/or have different average density.
Then, knowing  $\alpha$  (or $\beta$) and
the average asymmetry of a considered reaction product, 
 Eqs.(7) give 
the corresponding 
effective symmetry energy of the density region from which it emerges.
In other words, this analysis allows one  
to probe the local symmetry energy of clusterized systems. 
It should be noticed that this provides a different information
 with respect to the extraction of the total symmetry energy associated
with clusterized low-density matter \cite{francesca,Nato}. 

To conclude, in this paper we have undertaken a dynamical study of
the disassembly of two-component unstable systems, focusing on the coupling
between the development of isoscalar and isovector density fluctuations.
For nuclear systems, 
we have shown that the amplitude of isovector fluctuations follows the evolution 
of the local density and approaches, within
time scales compatible with nuclear reactions at Fermi energies,
the corresponding local equilibrium value, that is linked to the density-dependent 
symmetry free energy. 
Thus
fragment isospin fluctuations and isoscaling parameters are related 
to the symmetry energy at  
the fragment formation density.
These results are relevant to experimental isoscaling analyses aiming at 
extracting information on the symmetry energy, a topic of strong current interest in
nuclear physics and astrophysics \cite{Lattimer,betty,adri,betty1,dorso,trip,Ono}.
Though secondary decay effects are expected to reduce the 
sensitivity of these observables to the specific shape of the symmetry 
energy \cite{Ono},
this analysis should still allow one to probe the 
range of values spanned
within the low-density conditions reached in nuclear fragmentation reactions.
Finally, it should be noticed that our study
is perfomed within the semi-classical approximation.
It would be interesting to introduce
quantum fluctuations and investigate their influence 
on the relation between isoscaling,
isotopic distributions and symmetry energy.

{\bf Acknowledgments} - 
Illuminating discussions with F.Matera, F.Gulminelli and Ph.Chomaz are gratefully acknowledged.

 





\end{document}